\documentclass[ aps,%
floatfix,%
final,%
notitlepage,%
oneside,%
onecolumn,%
nobibnotes,%
nofootinbib,%
superscriptaddress,%
showpacs,%
]%
{revtex4}

\usepackage{epsfig}
\usepackage{amsfonts}
\usepackage{amsmath}
\usepackage{epsfig}
\usepackage{graphics}

\begin{document}

\title{Double charmonium production in exclusive bottomonia decays. }
\author{V.V. Braguta}

\email{braguta@mail.ru}

\author{A.K. Likhoded}

\email{Likhoded@ihep.ru}

\author{A.V. Luchinsky}

\email{Alexey.Luchinsky@ihep.ru}

\affiliation{Institute for High Energy Physics, Protvino, Russia}

\begin{abstract}
This paper is devoted to the leading twist exclusive bottomonia decays
with double charmonium in the final state. Using  models of the twist-2
charmonia distribution amplitudes the widths of these decays
are calculated within light cone formalism.
In addition, the processes under consideration are studied within
nonrelativistic QCD. In our analysis we have found that the production of some of the
$P$-wave charmonia mesons with $L_z \neq 0$ is allowed already at
the leading twist approximation. This means that the selection
rules which predict the suppression of such decays are violated.
The mechanism which lies behind this violation is discussed.
\end{abstract}

\pacs{
12.38.-t,  
12.38.Bx,  
13.66.Bc,  
}

\maketitle

\newcommand{\ins}[1]{\underline{#1}}
\newcommand{\subs}[2]{\underline{#2}}
\vspace*{-1.cm}
\section{Introduction.}

Double charmonium production at B-factories has been one of the most challenging
problem in quarkonium physics for many years. This problem appeared after
the measurements of the cross sections of the processes $e^+e^- \to J/\Psi \eta_c,
J/\Psi \eta_c', \psi' \eta_c, \psi' \eta_c', J/\Psi \chi_{c0}, \psi' \chi_{c0}$
at Belle \cite{Abe:2002rb, Abe:2004ww} and BaBar \cite{Aubert:2005tj} collaborations,
which were approximately by an order of magnitude larger than the leading order
nonrelativistic QCD (NRQCD) predictions \cite{Braaten:2002fi, Liu:1, Liu:2}.
There were many attempts
to resolve this discrepancy \cite{Bodwin:2002fk, Luchinsky:2003yh, Zhang:2005ch, Gong:2007db,
Zhang:2008gp, Ma:2004qf, Bondar:2004sv, Braguta:2005kr, Braguta:2006nf, Berezhnoy:2007sp,
Ebert:2008kj}. Lately, it was shown that the agreement between theory and experiment
can be achieved if one takes into account radiative  and relativistic  corrections to the
cross sections \cite{He:2007te, Bodwin:2007ga, Braguta:2008tg}.

Intensive study of double charmonium production at B-factories
has already led to the considerable theoretical progress in understanding of
hard exclusive processes with charmonia production. Further study of these processes
can improve our knowledge about QCD dynamic of the exclusive processes.  Moreover,
since double charmonium production is sensitive to the parameters
of charmonia wave functions, these processes can be used to study the structure
of charmonia mesons.

Another processes which can be used to study
the dynamic of hard exclusive processes and structure of charmonia mesons are
double charmonium production in bottomonia decays.
These processes are very similar to the processes of  double charmonia production
at B-factories since the masses of bottomonia are very near to the center mass energy
of $e^+e^-$ beams at B-factories $M_{b \bar b} \simeq \sqrt s = 10$ GeV.
At the same time exclusive bottomonia decays have very important advantage:
due to the different quantum numbers of the initial bottomonia one
has access to the final states whose production is suppressed at B-factories.

Although  exclusive bottomonia decays with double charmonium in the final state
are very interesting from theoretical point of view, thus far only
few papers were devoted to the study of some of these decays
\cite{Braguta:2005gw, Hao:2006nf, Jia:2007hy, Gong:2008ue}.
The present paper is devoted to the study of the leading twist double
charmonia production in bottomonia decays. To carry out this study
we are going to apply light cone formalism (LC) \cite{Lepage:1980fj, Chernyak:1983ej}.
Within this approach the amplitude of hard exclusive process
can be separated into two parts. The first part is partons production at very small
distances, which can be treated within perturbative QCD. The second part is
the hadronization of the partons at larger distances. For hard exclusive processes it can be
parameterized the by process independent distribution amplitudes (DA), which can be considered
as hadrons' wave functions at light like separation between the partons in the hadron.
It should be noted that DAs contain information about the structure of charmonia mesons.
To do the calculation of the decay widths  we are going to apply
models of the $S$- and $P$-wave charmonia DAs proposed in papers
\cite{Braguta:2006wr, Braguta:2007fh, Braguta:2007tq, Braguta:2008qe}.
We also apply NRQCD to study the same processes.

This paper is organized as follows. In the next section the description
of the approach applied in this paper is given. In this section
we show that the production of some $P$-wave charmonia mesons
with $L_z \neq 0$ is allowed already at the leading twist approximation,
what violates well known selection rules \cite{Chernyak:1983ej, Brodsky:1981kj}.
The mechanism which lies behind this violation is discussed.
In section III we present analytical expressions for the amplitudes
of the processes under study obtained within LC and NRQCD.
Numerical results of this paper are given in section IV.
In the last section we summarize the results of this paper.

\section{Description of the approach.}

\subsection{The amplitudes of the processes under study.}

In this paper  double charmonium production in exclusive bottomonia decays
will be considered. The presence of large energy scale $M_{b \bar b}$ which
is much greater than the masses of the final charmonia mesons
$M_{b \bar b} \gg M_{c \bar c}$ allows one to apply light cone formalism
\cite{Chernyak:1983ej} (LC). Within this formalism the amplitude of the
processes  $T$ is expanded  in inverse powers of the large energy scale
$1/M_{b \bar b}$
\begin{eqnarray}
T = \frac {t_0} { M_{b \bar b}^n} + \frac {t_1} { M_{b \bar b}^{n+1}} + ...
\end{eqnarray}
In this paper we are going to consider the bottomonia decays which are different
from zero already at the leading approximation in $1/M_{b \bar b}$ expansion. From here on these
processes will be called the leading twist processes. Moreover,
the calculation done in this paper will be restricted by the calculation
of the first nonvanishing contribution in $1/M_{b \bar b}$ expansion.
At this level of accuracy the amplitude of hard exclusive production
of mesons $M_1, M_2$ in bottomonia decay can be written in the following from
\begin{eqnarray}
T=\int_{-1}^{1} \int_{-1}^{1} d \xi_1 d \xi_2 H(\xi_1, \xi_2) \phi_{M_1} (\xi_1) \phi_{M_2} (\xi_2),
\label{amp}
\end{eqnarray}
where $\xi_1, \xi_2$ are the fractions of the relative momenta of the whole meson carried
by the quark-antiquark pair in the mesons $M_1, M_2$ correspondingly, $H(\xi_1,\xi_2)$
is the hard part of the amplitude, $\phi_{M_1} (\xi_1), \phi_{M_2} (\xi_2)$ are the
leading twist distribution amplitudes (DA) of the mesons $M_1, M_2$.

Now few comments are in order.

{\bf 1.} The leading twist DAs $\phi_{M_1} (\xi_1), \phi_{M_2} (\xi_2)$ parameterize infinite
series of the twist-2 operators. For instance, the leading twist DA of the pseudoscalar meson
$P$ parameterizes the operators $\bar Q \gamma_5 \gamma_{+} D_{+}^n Q,~~n=0,1,2..$ as
follows \footnote{Here the designation $a_+=a_{0}+a_{z}$ for fourvector $a_{\mu}$ is used . It is also assumed that
very energetic meson $P$ moves along $z$ direction.}
\begin{eqnarray}
\langle P(q) | \bar Q  \gamma_5 \gamma_+
D_{+}^n Q | 0 \rangle = i f_P (q_+)^{n+1} \int_{0}^1  dx \phi(\xi) \xi^n,
\label{wf}
\end{eqnarray}
where $q$ is the momentum of the pseudoscalar meson $P$, $f_P$ is the constant which is defined as
$\langle P(q) | \bar Q \gamma_{\mu} \gamma_5 Q |0 \rangle = i f_P q_{\mu}$. One can also
think of $\phi(\xi)$ as about the amplitude to find the quark-antiquark pair
in the meson $P$ with the fraction of the relative momentum of the whole meson $\xi$.
It should be noted that DAs parameterize the nonperturbative effects in the amplitude.
In Appendix A we collect the definitions of all leading twist charmonia DAs needed
in this paper.

From formulas (\ref{amp}), (\ref{wf}) one sees that amplitude (\ref{wf})
resums infinite series of the twist-2 operators in the amplitude. If the mesons $M_1, M_2$
are  nonrelativistic mesons, formula (\ref{amp}) resums relativistic corrections to
the amplitude $T$.

{\bf 2.} The hard part of the amplitude $H(\xi_1, \xi_2)$ describes small distance effects,
which can be calculated within perturbative QCD. At the same time the DAs $\phi_{M_1} (\xi_1), \phi_{M_2} (\xi_2)$ parameterize nonperturbative effects,
which take place at large distances. From this one can conclude that
formula (\ref{amp}) separates the effects of small and large distances.

In this paper we study the processes with bottomonia mesons in the initial states.
Below the bottomonia mesons will be described at the leading order approximation of NRQCD.
At this level of accuracy the amplitudes of $P$-wave bottomonia decays
contain the contributions coming from color-octet states \cite{Bodwin:1994jh}.
However, it is clear that the contributions of such states to the amplitude of
hard exclusive decays are suppressed.
This can be seen as follows. Besides
 the quark-antiquark pair, color-octet state contains one additional gluon.
The amplitude to attach this gluon to the one of the outgoing charmonia is
strongly suppressed
since this gluon does not have enough energy. So, this gluon
must be absorbed in the hard part of the amplitude. What leads to the
suppression in $\alpha_s$ and higher powers of relative velocity of bottomonia.
For this reason, only color singlet states will be taken into the account in the
present analysis.
The calculation will be done using the technique of  the projection operators
\cite{Bodwin:2002hg, Braaten:2002fi}.

As it was noted the hard part of the amplitude $H(\xi_1, \xi_2)$ contains the small distance effects.
At small distances the strong coupling constant $\alpha_s$ is small, so one can expand the
$H(\xi_1, \xi_2)$ in a series over $\alpha_s$. However, presence of two strongly separated energy scale $M_{b \bar b} \gg M_{c \bar c}$, gives rise to the appearance of large logarithm
$\log {E_h^2/M_P^2}\sim \log(M_{bb}^2/M_{cc}^2)$. This logarithm enhances the role of radiative corrections. The main contribution
to amplitude (\ref{amp}) comes from the leading logarithmic radiative corrections
$\sim (\alpha_s \log {E_h^2/M_P^2})^n$. It turns out that these corrections can be taken
into the account in formula (\ref{amp}) if this formula is rewritten as follows \cite{Chernyak:1983ej, Lepage:1980fj}
\begin{eqnarray}
T=\int_{-1}^{1} \int_{-1}^{1} d \xi_1 d \xi_2 H(\xi_1, \xi_2, \mu) \phi_{M_1} (\xi_1, \mu) \phi_{M_2} (\xi_2, \mu),
\label{amp1}
\end{eqnarray}
To resum the leading logarithmic radiative corrections coming from all loops the scale $\mu$ should
be taken of order of $\sim M_{b \bar b}$. The hard part of the amplitude $H(\xi_1, \xi_2, \mu)$
should be calculated at the tree level approximation. At this level $H(\xi_1, \xi_2, \mu)$
depends on the renormalization scale $\mu$ only through the running of the strong coupling
constant $\alpha_s(\mu)$. The rest of the leading logarithms are resummed in the
DAs $\phi_{M_1} (\xi_1, \mu) \phi_{M_2} (\xi_2, \mu)$ using renormalization group
method. It should be stressed that formula (\ref{amp1}) exactly resums the leading logarithmic
radiative corrections which appear in all loops. In the calculation we use $\mu=M_{b \bar b} / 2$.

Commonly, to study the production of nonrelativistic mesons one uses
effective theory NRQCD \cite{Bodwin:1994jh}. NRQCD deals with three energy scales $m_c \gg m_c v \gg m_c v^2$,
where $m_c$ is the mass of c-quark, $v\ll 1$ is relative velocity of quark antiquark pair in charmonium.
In the process of hard nonrelativistic meson production there appears one additional energy scale
$E_h \sim M_{b \bar b}$ which is much greater than all scales $m_c, m_c v, m_c v^2$. Evidently, it is not
possible to apply NRQCD at this scale. From the effective theory perspective,
first, this large energy scale must be integrated out. And this is done through
the taking into account renormalization group evolution of the DAs
$\phi_{M_1} (\xi_1, \mu) \phi_{M_2} (\xi_2, \mu)$.

\subsection{Selection rules.}

In this section we are going to determine what processes
of double charmonium production in bottomonia decays are
the leading twist processes. To do this one should
determine the asymptotic behaviour of the amplitudes of the
processes under study in the limit $M_{b \bar b} \to \infty$.
The determination of the asymptotic behaviour was a subject
of many papers \cite{Chernyak:1983ej, Chernyak:1977fk,Chernyak2,
Chernyak3, Brodsky:1981kj}. It turns out that this behavior is
determined by the quantum numbers of the final hadrons and does not
depend on the DAs of the final hadrons. Below we are going to follow paper
\cite{Chernyak:1983ej}. The authors of this paper formulated general
selection rules which can be used to specify the leading twist
processes

{\bf 1. } For a very energetic charmonium the $c$-quark helicity coincides with the projection of its
spin into the direction of the hadron momentum. This rule is valid up to the
corrections $\sim k_{\perp}/M_{b \bar b}$, where $k_{\perp}$ is the transverse
momentum of the quark in charmonium.

{\bf 2. } For the charmonia states with $L_z \neq 0$ ($L$ is a quark angular momentum,
the hadron is assumed to move along $z$-axis) the asymptotic behavior of the amplitude
is power suppressed.

{\bf 3. } The QCD interaction has a vector nature and at the quark-gluon vertex
the quark helicity is conserved (up to the corrections $\sim m_c/M_{b \bar b}$).

At the leading order approximation in the $\alpha_s$, the exclusive decays of the
C-even bottomonia can be described by the Feynman diagrams similar to that shown in
Fig. 1a. Typical diagram for the C-odd bottomonia exclusive decays is shown in Fig. 1b.
Applying selection rules (1)-(3) and diagrams shown in Fig. 1a,b one can
prove that there are two types of the leading twist processes. The first one contains the processes
in which the spin of the initial state $J_z=0$ ($z$-axes is chosen in the direction of motion of final
charmonia mesons). In this case the helicities of final mesons are $\lambda_1=\lambda_2=0$.
Any bottomonium meson can decay in this way. The second type of the processes contains
the processes with $J_z=\pm 2$ and $\lambda_1=-\lambda_2=\pm 1$. Evidently, only
the $\chi_{b2}$ meson can satisfy this condition. Note also, that one can introduce
the quantity $\Lambda=\lambda_1+\lambda_2$ and for all leading twist processes
this quantity is zero. The amplitude of the process with $\Lambda \neq 0$
is suppressed as $\sim (M_{c \bar c }/M_{b \bar b })^{\Lambda}$ relatively to the
$\Lambda=0$ process.

\begin{figure}
\begin{centering}
\includegraphics{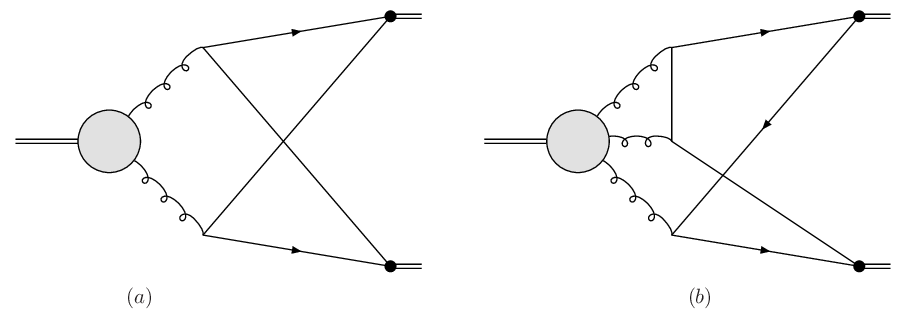}
\par\end{centering}

\caption{Typical diagrams for $C$-even (fig. a) and $C$-odd (fig. b) bottomonia
decays into charmonia pair\label{fig:diags1}}

\end{figure}

For the bottomonia with $J_z=0$ one can derive one more selection
rule. To do this let us introduce the quantum number "naturalness": $\sigma=(-1)^S P$,
where $P$ is the parity, $S$ is the spin. If the naturalness is not conserved:
$\sigma_{initial} \neq (\sigma_1 \sigma_2)_{final}$, there appears the antisymmetric
tensor $\epsilon_{\mu \nu \rho \sigma}$ in the amplitude. This tensor is contracted
with the polarizations and momenta of the final charmonia.
However, the polarization vector of energetic longitudinally polarized meson is proportional to its momentum, so
for the
mesons with the longitudinal polarizations there will be no enough fourvectors
to get nonzero result after the contraction. From this one can conclude that
for the leading twist processes the naturalness is conserved.

Applying the rules that were presented  above one can find all leading twist
exclusive bottomonia decays. Some of these decays are the decays of the C-even
bottomonia mesons and some are the decays of the C-odd bottomonia mesons. Evidently  the
C-odd bottomonia decays can proceed through at least tree gluons (see Fig. 1b).
From this one can conclude that the amplitudes of the C-odd bottomonia decays
are suppressed as $\sim \alpha_s/\pi \sim 0.07$ in comparison with the C-even
bottomonia decays. For this reason, we are not going to consider these
decays. So,  below the following decays will be considered
\begin{eqnarray}
\eta_{b},~ \chi_{b1} &\to&  h_{c}~ J/\psi (\psi'),~~
\eta_{c} (\eta_c')~ \chi_{c0}(\chi_{c2}),~~\chi_{c1}~\chi_{c0}(\chi_{c2}), \nonumber  \\
\chi_{b0},~ \chi_{b2}  &\to&  \eta_{c} (\eta_{c}')~\chi_{c1},~~\eta_{c}(\eta_{c}')~ \eta_{c}(\eta_{c}'),
~~ J/\psi(\psi')~ J/\psi(\psi'),~~h_{c}~h_{c},~~\chi_{c0}~\chi_{c2},~~\chi_{c0}~\chi_{c0}, \nonumber \\
&& \chi_{c1}~\chi_{c1},~~\chi_{c2}~\chi_{c2}
\nonumber \\
\chi_{b2} & \to & h_{c}J/\psi (\psi'),\,\chi_{c1}\chi_{c2}.
\label{pros}
\end{eqnarray}

\subsection{Violation of the selection rules.}

Let us consider a longitudinally polarized $\chi_{c1}$ meson moving along $z$-axes.
In this meson the spin $S=1$ and orbital momentum of $L=1$ of the quark-antiquark pair
sum to $J=1$ state. The Clebsch-Gordan coefficients \cite{Yao:2006px} of the longitudinally polarized
$\chi_{c1}$ meson ($J_z=0$) are $C^{J_z=0}_{L_z=+1, S_z=-1}=-C^{J_z=0}_{L_z=-1, S_z=+1}=1/\sqrt 2,
C^{J_z=0}_{L_z=0, S_z=0}=0$. This means that the longitudinal polarization can be realized
only through the $L_{z}=\pm 1$. However, the selection rules tell us that
such states are suppressed. From this one can draw a conclusion: the processes
with the production of the longitudinally polarized $\chi_{c1}$ meson
(for instance, the $\chi_{b0} \to \eta_c \chi_{c1}$ decay)
are not the leading twist processes.
Similar situation takes place for a transversely polarized
$h_c$ meson. In this case one has $S=0, L=1$. Evidently, the state $J_z=\pm 1$ can appear
only if $L_z=\pm 1$. So, from the selection rules one can conclude that the production
of the transversely polarized $h_c$ meson is suppressed. Nevertheless, the calculation
done within NRQCD shows that these conclusions are not correct. The processes shown in list (\ref{pros}) with the
production of the transversely polarized $h_c$ meson and longitudinally polarized $\chi_{c1}$
meson are the leading twist processes.

To demonstrate that the production of the  longitudinally
polarized $\chi_{c1}$ meson is not suppressed, let us consider the process of
the production of the $\eta_c$ meson and $\{ \bar c c \}$ pair $\chi_{b0} \to \eta_c \{ \bar c c \}$.
Assume further that $\{ \bar c c \}$ pair has very small invariant mass
$M_{\bar c  c} \ll M_{\bar b b}$. So, we deal with quasi-exclusive process for
which one can apply LC technique. For instance, one can expand the amplitude of this process
in $1/M_{\bar b b}$ and use the selection rules. Similarly to  exclusive processes,
the leading order contribution in $1/M_{\bar b b}$ expansion for quasi-exclusive process appears due to the leading twist
operator of the $\eta_c$ meson. Feynman diagrams that give contribution to
this process are similar to that shown in Fig. 1a. It is not difficult to see that the
leading order contribution to the amplitude in $1/M_{\bar b b}$ expansion
can be written as follows
\begin{eqnarray}
T=B^{\mu \nu} \times \bigl \{  \bar c  \gamma_{\mu} \hat P \gamma_{\nu} c \bigr \},
\end{eqnarray}
where $B^{\mu \nu}$ is proportional to the amplitude that describes the decay of the $\chi_{b0}$
into two gluons, $\hat P = \hat p_1 \gamma_5$ is the dirac structure of the leading twist wave function of the
$\eta_c$ meson with momentum $p_1$.  The tensor $B^{\mu \nu}$ can depend only on the
tensor structures $q_1^{\mu} q_2^{\nu}, q_2^{\mu} q_1^{\nu},
q_2^{\mu} q_2^{\nu}, q_1^{\mu} q_1^{\nu}, (M_{\bar b b})  g^{\mu \nu}$. Evidently,
the leading twist contribution to the amplitude originates only from the last
tensor structure. The others give terms proportional to masses. From this one can conclude
that the leading twist contribution to the amplitude $T$ is proportional to
\begin{eqnarray}
T \sim \bar c  \gamma_{\mu} \hat p_1 \gamma_5 \gamma^{\mu} c  \sim
\bar c  \hat p_1 \gamma_5  c .
\end{eqnarray}
Now let us consider what $\bar c c$ states contribute to the amplitude $T$.
To do this let us consider the expression for $T$ in the center mass of $\bar c c$
pair and assume that $\bar c c$ is nonrelativistic pair. In this case one can expand
the operator $\bar c  \gamma_{\mu} \gamma_5  c$ in relative velocity of quark-antiquark
pair using Foldy-Wouthuysen-Tani transformation \cite{Foldy:1949wa}. The terms relevant
to the production can be written as follows
\begin{eqnarray}
\bar c  \gamma_{\mu} \gamma_5  c= \delta^{\mu 0}~
\psi^+ \biggl [ 1 + \frac {\bf D^2} {2 m_c^2}  \biggr ]  \chi +
\delta^{\mu i}~\frac 1 {m_c} \psi^+ \biggl [ {\bf D \times \sigma }  \biggr ]_i  \chi + O(v^3),
\label{ex}
\end{eqnarray}
where $\bf D$ is vector part of the gauge derivative,
the index $i$ is vector index, $m_c$ is the mass of $c$-quark,
$\psi^+$ and $\chi$ are Pauli spinor fields that create a quark
and an antiquark respectively. Now it is clear that the operator
$\psi^+ \biggl [ 1 + \frac {\bf D^2} {2 m_c^2}  \biggr ]  \chi$
corresponds to the $L=0, S=0, J=0$ state. So, one can recognize $\eta_c$ meson
contribution in this operator. From this one can conclude that
the process $\chi_{b0} \to \eta_c \eta_c$ is the leading twist process.
The operator $\psi^+ \biggl [ {\bf D \times \sigma }  \biggr ]  \chi$ corresponds
to the $L=1, S=1, J=1$ state. So, one sees that this operator creates
the longitudinal $\chi_{c1}$ meson already at the leading twist approximation
what violates rule (2) of the selection rules. Similarly, one can show that
the transversely polarized $h_c$ meson can be produced at the leading
twist approximation.

Now, let us try to understand why the second selection rule is violated?
To answer this question first let us recall how this selection rule originates.
In LC leading twist distribution amplitude $\phi(x)$ of nonrelativistic
meson $M$ moving along $z$-axes can be written in the following form
\begin{eqnarray}
\phi(x,Q^2) \sim \int^{k_{\perp}^2<Q^2} d^2 k_{\perp} \psi_M (k_{\perp},x),
\label{intDA}
\end{eqnarray}
where $\psi_M(k_{\perp},x)$ is the hadron wave function of the mesons $M$, $x$ is the
fraction of momentum carried by quark. The meson $M$ with $L_z \neq 0$
has the function $ \psi_M (k_{\perp},x)$ which contains the factor $\sim \exp {(i L_z \varphi)}$.
So, the integration over $d^2 k_{\perp}$ acts as a projector to
the $L_{z}=0$ state and $L_z \neq 0$ states cannot appear at the
leading twist approximation. Similar arguments were used in papers
\cite{Chernyak:1983ej, Brodsky:1981kj} to prove the second selection rule.

Further, this point will be considered in more detail. To get $L_z \neq 0$ in addition
to the $\exp {(i L_z \varphi)}$ one should find somewhere the factor $k_{\perp}$.
It can be found in the matrix element of hard exclusive process or
in the expansion of the bispinors of quark-antiquark pair in relative
momentum (see formula (\ref{ex}) ). The first source of the $k_{\perp}$
is not very important since it always leads to the suppression of the
amplitude, so it is beyond the leading twist accuracy.

Consider the case when the $k_{\perp}$ appears due to the expansion
of quark-antiquark bispinors in relative momentum. Formula (\ref{ex}) proposes
the example of the expansion of the operator $\bar c \gamma_{\mu} \gamma_5 c$
in relative momentum. As it was already noted, the first term of this expansion
corresponds to the leading twist operator of the $\eta_c$ meson. This can be
seen from the fact that the leading twist operators among the operators
of the fixed dimensionality are those which are maximally enhanced due to the
Lorentz transformation from the center mass frame to the frame where the meson has
very large energy. Evidently, the $\delta_{\mu 0}$ in the center mass frame
can be rewritten in a covariant way: $\delta_{\mu 0} = p_{\mu}/M$, where $p$ and $M$ are the
momentum and mass of the meson. From this one can conclude that  due to the Lorentz transformation the first
operator in formula (\ref{ex}) will be enhanced by the first power
of large $\gamma$-factor, what exactly corresponds to the leading twist (see list
of the leading twist DAs in Appendix A).

The second term in formula (\ref{ex}) corresponds to the longitudinally
polarized $\chi_{c1}$ meson. This term contains gauge derivative which becomes
$\psi^+ D_{\perp} \chi \sim k_{\perp}$. So, due to this term one has additional
$k_{\perp}$ which leads to the nonzero contribution of the $L_z \neq 0$
states. Here few comments are in order. First, note that in the bispinors
the term $\sim k_{\perp}$ goes with Pauli matrices $\sigma_{\perp}$ which
flip the spin of quark or antiquark. This leads to the violation
of the first selection rule. So, the second term leads to the violation
of the first and second selection rules. Second, it should be noted
that the states $(S=0, S_z=0)$ and $(S=1, S_z=\pm 1)$ cannot
be produced simultaneously without one spin flip. This spin flip can be
realized due to the violation of the first rule or the third rule.
The violation of the third rule always put the process beyond the
leading twist accuracy, since it is based on the vector nature
of the quark-gluon vertex. So, the violation of the first selection
rule in the wave function is needed in order to avoid the violation
of the third selection rule.

Now let us do Lorentz transformation of the second term in
formula (\ref{ex}) from the center mass frame to the frame
where the meson has very large energy. From the first
sight, transverse components of fourvectors will not be
enhanced by $\gamma$-factor. This means that the violation
of the first and second selection rules leads to the
appearance of the transverse components of fourvectors in the
operators what puts the amplitude of such process beyond the
leading twist accuracy. This conclusion is in agreement
with the results of papers \cite{Chernyak:1983ej, Brodsky:1981kj}.
However, note that the $k_{\perp}$ and $\sigma_{\perp}$
appear in formula (\ref{ex}) in the form of vector product.
This means that covariant expression of formula (\ref{ex})
has the form
\begin{eqnarray}
\bar c  \gamma_{\mu} \gamma_5  c= \frac {p_{\mu}} {M}~
\psi^+ \biggl [ 1 + \frac {\bf D^2} {2 m_c^2}  \biggr ]  \chi + \frac {1} {m_c}
e_{\nu \mu \lambda \rho} \frac {p_{\nu}} {M} \psi^+ \biggl [ {\bf D^{\lambda}  \sigma^{\rho} }  \biggr ]  \chi + O(v^3).
\end{eqnarray}
It is seen from this equation that although transverse components
are not enhanced the structure $e_{\nu \mu \lambda \rho} {p_{\nu}}$
compensates this drawback and returns the second term to the set
of the leading twist operators.

At the end of this section we would like to note that
the violation of the selection rules in the case of the transversely
polarized $h_c$ meson is similar to that of the longitudinally polarized $\chi_{c1}$ meson.
So, the conclusion of this section is {\it the production of the
longitudinally polarized $\chi_{c1}$ meson and transversely
polarized $h_c$ meson is nonzero at the leading twist approximation due to
the appearance of the special structures in the corresponding DAs which compensate
the penalty for the violation of the first and second selection rules.}

\subsection{Description of the calculation procedure.}

To calculate  processes (\ref{pros}) at the leading twist approximation of
LC one can apply the following rules. As it was noted above, the bottomonia
mesons will be treated at the leading order approximation of NRQCD
using the technique of  the projection operators \cite{Bodwin:2002hg, Braaten:2002fi}.
Applying this technique one can calculate the amplitude of the decay
of some bottomonium into two quark-antiquark pairs. The total momentum and relative
momentum of the first pair are $p_1$ and $\xi_1 p_1$. The total momentum and relative
momentum of the second pair are $p_2$ and $\xi_2 p_2$. At large distances
these pairs become $M_1$ and $M_2$ charmonia mesons with the momenta $p_1$, $p_2$.
\footnote{It should be noted that at the leading twist approximation one can
disregard the masses of quarks and hadrons.}
In LC the hadronization of quark-antiquark pair is described by  DAs. To
calculate the amplitude of meson production one should replace bispinors
of quark-antiquark pair $v \bar u$ in the amplitude by the corresponding
distribution amplitude. The list of the leading twist DAs needed
in the calculation can be found in Appendix A. After taking the traces over
dirac and color indexes one can get analytical expression for the amplitude.

Another independent approach that can be used to calculate the amplitude of double charmonium
production in bottomonia decays is NRQCD. In this approach the final charmonia
mesons are treated as nonrelativistic states. In this paper we are going
to apply NRQCD at the leading order approximation in relative velocity in
charmonia. It should be noted that this approximation cannot be considered
as reliable. This conclusion can be drawn from the experience obtained in
the study of double charmonium production at B-factories \cite{Braguta:2008tg}, which tells us that
relativistic and radiative corrections to double charmonia production can be
very large. Nevertheless, in this paper the leading order approximation of NRQCD
will be used to get independent estimation of the widths of the exclusive
bottomonia decays under study.

In addition to the estimation of the widths of the bottomonia decays, NRQCD can be
used to study very important question of LC: the estimation of the corrections to the
leading twist approximation (power corrections). To estimate power corrections
one can apply the idea of duality of NRQCD and LC descriptions
of the hard exclusive nonrelativistic mesons production: {\it if the amplitudes
of the hard process under study obtained within NRQCD and LC are expanded
in powers of inverse hard energy scales and relative velocity one will get
series equal to each other. } It should be noted that these
is no strict proof of this statement. Moreover, the general proof for
the NRQCD factorization of exclusive process involving P-wave heavy
quarkonium is not available (see, e.g., \cite{Bodwin:2008nf}). There may be large
soft contributions at order $m_c/m_b$ to these decays. However, one can expect that this statement is
 true since the amplitude in NRQCD and LC can be expended in  series of equivalent
operators. Assuming that both theories can describe experiment one can expect that
these expansions in both theories coincide. Note also that we have checked that
all expressions for the amplitudes under study obtained in LC and expanded in relative
velocity coincide with that obtained in NRQCD and expanded in $1/M_{b \bar b}$.

Now it is clear how it is possible to estimate the size of power corrections in LC.
To do this one should take the leading order NRQCD prediction for the amplitude and
expand it in $M_{c \bar c}/M_{b \bar b} \simeq m_c/m_b$. The first term of this
expansion can be reproduced by the leading twist approximation of LC expanded in relative
velocity in charmonia. The second term in this expansion can be reproduced by
the power corrections to the leading twist contribution. So, the second term
of NRQCD expansion in $\sim m_c/m_b$ gives the estimation of power correction
to the leading twist approximation. Below this approach will be used to estimate
the error of the calculation.

\section{Analytical expressions for the matrix elements.}

In this section analytical expressions for the amplitudes and the widths
of the double charmonium production in bottomonia decays obtained
within the leading order approximation of LC and NRQCD are
given. Before the analytical expressions are given, let us
introduce some useful designations.
First, we introduce the definitions of the constants of the vector ($V=J/\Psi, \psi'$)
and pseudoscalar ($P=\eta_c, \eta_c'$) mesons that parameterize
the following matrix elements
\begin{eqnarray}
\langle V (p, \epsilon) | \bar C \gamma_{\alpha} C| 0 \rangle &=& f_{V}^L M_{V} \epsilon_{\alpha}
\nonumber \\
\langle V (p, \epsilon) | \bar C \sigma_{\alpha \beta} C| 0 \rangle_{\mu} &=&  f_{V}^T (\mu)
(p_{\alpha} \epsilon_{\beta} - p_{\beta} \epsilon_{\alpha} )
 \nonumber \\
\langle P (p) | \bar C \gamma_{\alpha} \gamma_5 C| 0 \rangle &=& i f_{P}^L p_{\alpha}.
\label{constS}
\end{eqnarray}
Next, we introduce the following constants for the $P$-wave charmonia
mesons $\chi_{c0}, \chi_{c1}, \chi_{c2}, h_c$
\begin{eqnarray}
\langle \chi_{c0} (p) | \bar C  \gamma_{\alpha} (-i  {\overset {\leftrightarrow} {D}}_{\nu} )  C| 0 \rangle
&=& f_{\chi{0}}^L (\mu) \bigr ( p_{\alpha} p_{\nu} - M^2_{\chi_{c0}} g_{\alpha \nu}  \bigl ) , \nonumber \\
\langle \chi_{c1} (p, \epsilon) | \bar C  \gamma_{\alpha} \gamma_5  C| 0 \rangle
&=& f_{\chi{1}}^L  M_{\chi_{c1}} \epsilon_{\alpha},
~~~~
\langle \chi_{c1} (p, \epsilon) | \bar C  \sigma_{\alpha \beta} (-i  {\overset {\leftrightarrow} {D}}_{\nu} )
  C| 0 \rangle = f_{\chi{1}}^T(\mu) e_{\alpha \beta \rho \lambda} \epsilon^{\rho} p^{\lambda} p_{\nu},
\nonumber \\
\langle h_c (p, \epsilon) | \bar C  \gamma_{\alpha} \gamma_5
(-i  {\overset {\leftrightarrow} {D}}_{\nu} )  C| 0 \rangle &=&
f_{h}^L(\mu) M_{h_c} p_{\alpha} \epsilon_{\nu} ,
~~~~~~~~~~
\langle h_{c} (p, \epsilon) | \bar C  \sigma_{\alpha \beta}   C| 0 \rangle
= f_{h}^T(\mu)  e_{\alpha \beta \rho \lambda} \epsilon^{\rho} p^{\lambda}, \nonumber \\
\langle \chi_{c2} (p, \epsilon) | \bar C  \gamma_{\alpha}
(-i  {\overset {\leftrightarrow} {D}}_{\nu} )  C| 0 \rangle &=&
f_{\chi{2}}^L(\mu) M_{\chi_{c2}}^2  \epsilon_{\alpha \nu} ,
~~~~~~~~~~
\langle \chi_{c2} (p, \epsilon) | \bar C  \sigma_{\alpha \beta} ( -i  {\overset {\leftrightarrow} {D}}_{\nu} )  C| 0 \rangle
= i f_{ \chi{2} }^T(\mu) ( p_{\alpha} \epsilon_{\beta \nu} - p_{\beta} \epsilon_{\alpha \nu} )
\label{constP}
\end{eqnarray}
It should be noted here that for the mesons $\chi_{c1}, h_c$ the polarization $\epsilon$
is described by the four vector $\epsilon_{\mu}$, for the $\chi_{c2}$ meson the
polarization $\epsilon$ is described by the tensor $\epsilon_{\mu \nu}$.
The superscript $T$ in formulas (\ref{constS}), (\ref{constP}) means that
the corresponding meson is transversely polarized ($\lambda=\pm 1$).
The superscript $L$ implies that the meson has helicity $\lambda=0$.
Except the constants $f_V^L, f_P, f_{\chi1}^L$, all constants in formulas (\ref{constS}), (\ref{constP})
are scale ($\mu$) dependent quantities. The anomalous dimensions of these constants can be
found in papers \cite{Braguta:2007fh, Braguta:2008qe}.

In addition to  constants (\ref{constS}), (\ref{constP}), one needs the bottomonia
NRQCD matrix elements $\langle O_S^{b \bar b} \rangle$, $\langle O_P^{b \bar b} \rangle$
which are defined in \cite{Bodwin:1994jh}. The matrix element $\langle O_S^{b \bar b} \rangle$
is proportional to the S-wave bottomonium radial wave function at the origin.
The matrix element $\langle O_P^{b \bar b} \rangle$
is proportional to the of the derivative of P-wave bottomonium radial wave function
at the origin.


{\bf The decays of the $\eta_b$ meson.} The decays $\eta_{b} \to  h_{c}~ J/\psi (\psi'),~~
\eta_{c} (\eta_c')~ \chi_{c0}(\chi_{c2}),~~\chi_{c1}~\chi_{c0}(\chi_{c2})$ are the leading twist decays of
the $\eta_b$ meson. According to the discussion in the previous section at the leading twist approximation
of LC only the production of the longitudinally polarized mesons is allowed. The same is
true for the decays of all bottomonia except the $\chi_{b2}$ meson. Corresponding
amplitude for all the leading twist decays of the $\eta_b$ meson can be written as
\begin{eqnarray}
\mathcal{M}(\eta_{b}\to M_{1}M_{2}) & = &
\frac{64\pi^{2}\alpha_{s}^{2} }{27}\sqrt{\frac{\langle O_{S}^{b \bar b} \rangle }{m_{b}}}\frac{f_{1}^{L}f_{2}^{L}}{m_{b}^{2}}I_{0}^{(\eta_{b})},
\label{ampl1}
\end{eqnarray}
 where $f_{1,2}^{L}$ are the mesonic constants of longitudinally polarized
charmonia and
\begin{eqnarray*}
I_{0}^{(\eta_{b})} & = & \int\limits _{-1}^{1}d\xi_{1}d\xi_{2}\frac{\xi_{1}+\xi_{2}}{(1-\xi_{1}^{2})(1-\xi_{2}^{2})(1+\xi_{1}\xi_{2})}\phi_{1}^{L}(\xi_{1})\phi_{2}^{L}(\xi_{2}),
\end{eqnarray*}
where $\phi_{1}^{L}(\xi_{1})\phi_{2}^{L}(\xi_{2})$ are the final charmonia DAs.
It is not difficult to get the amplitude of any leading twist decay of the $\eta_b$
meson using formula (\ref{ampl1}). For instance, to get the amplitude of the process
$\eta_b \to \eta_c \chi_{c2}$ one should take $f_1^L=f_P^L, f_2^L=f_{\chi2}^L$ and
the longitudinal DAs (see Appendix A) of the corresponding mesons.

Using this matrix element it is easy to obtain the expression for the $\eta_b\to M_1 M_2$ decay width within light cone formalism:
\begin{eqnarray}
\Gamma\left( \eta_b \to M_1 M_2\right) &=&
    \frac{128\pi^3 \alpha_s^4\beta}{729} \frac{\langle O_S^{bb}\rangle}{m_b^2}\left[
        \frac{f_l^L f_2^L}{m_b^2}  I_0^{(\eta_b)}
    \right]^2,
\end{eqnarray}
where
\begin{eqnarray*}
  \beta = \frac {2 |p_f|} {M} = \frac{\sqrt{M^2-(m_1+m_2)^2}\sqrt{M^2-(m_1-m_2)^2}}{M^2}, 
\end{eqnarray*}
is the factor which results from the phase space integration, $p_f$ is the momentum of final meson, 
$M$ and $m_{1,2}$ are masses of bottominium and chamonium mesons respectively.
Within NRQCD one can get the following expressions for the width of the processes
under study
\begin{eqnarray*}
\Gamma(\eta_b\to M_1 M_2) &=& \frac{128\pi^3\alpha_s^4\beta}{729}\frac{\langle O_{S}^{b \bar b} \rangle}{m_b^2}
  \left[\frac{f_1^\mathrm{NRQCD} f_2^\mathrm{NRQCD}}{m_b^2}\right]^2 F(\eta_b \to M_1 M_2),
\end{eqnarray*}
where
\begin{eqnarray}
f_i^\mathrm{NRQCD} &=& f_S^\mathrm{NRQCD}= \sqrt{\frac{\left<O_S^{cc}\right>}{m_c}}
\label{fsnr}
\end{eqnarray}
for $S$-wave charmonium states and
\begin{eqnarray}
f_i^\mathrm{NRQCD} &=& f_P^\mathrm{NRQCD}=\sqrt{\frac{\left<O_P^{cc}\right>}{m_c^3}}
\label{fpnr}
\end{eqnarray}
for $P$-wave charmonium states. The factor $F[\eta_b \to M_1 M_2]$ for different final states is
\begin{eqnarray*}
F[\eta_b \to h_c \psi] &=& 1+32 r^2,\\
F[\eta_b \to \eta_c \chi_{c0}] &=& \frac{1}{3}+\frac{16 r}{3}+\frac{64 r^2}{3},\\
F[\eta_b \to \eta_c \chi_{c2}] &=& \frac{2}{3}-\frac{16 r}{3}+\frac{32 r^2}{3},\\
F[\eta_b \to \chi_{c0} \chi_{c1}] &=& \frac{2}{3}+\frac{32 r}{3}+\frac{40 r^2}{3}-\frac{800 r^3}{3},\\
F[\eta_b \to \chi_{c2} \chi_{c1}] &=& \frac{4}{3}-\frac{32 r}{3}+\frac{104 r^2}{3}-\frac{160 r^3}{3}
\end{eqnarray*}
where $r=m_c/m_b$.

It should be noted here that in any exclusive bottomonium decay within LC the hard 
scale is set up by the mass of the bottomonium meson $M$, but not by the double pole mass 
of $b$-quark $2 m_b$. Within NRQCD both approaches are equivalent at the leading order approximation.   
For this reason, in the calculation it is assumed that $m_b=M/2$.

{\bf The decays of the $\chi_{b0}$ meson.}
The decays $\chi_{b0}  \to  \eta_{c} (\eta_{c}')~\chi_{c1},~~\eta_{c}(\eta_{c}')~ \eta_{c}(\eta_{c}'),
~~ J/\psi(\psi')~ J/\psi(\psi'),~~h_{c}~h_{c},~~\chi_{c0}~\chi_{c2},
~~\chi_{c0}~\chi_{c0},\\ \chi_{c1}~\chi_{c1},~~\chi_{c2}~\chi_{c2}$
are the leading twist decays of the $\chi_{b0}$ meson.
The amplitude for these decays can be written as
\begin{eqnarray*}
\mathcal{M}\left(\chi_{b0}\to M_{1}M_{2}\right) & = & -\frac{256\pi^{2}\alpha_{s}^{2}}{27\sqrt{3}}\sqrt{\frac{\left\langle O^{b\bar b}_{1}\right\rangle _{P}}{m_{b}^{3}}}\frac{f_{1}^{L}f_{2}^{L}}{m_{b}^{2}}I_{0}^{(\chi_{b0})},
\end{eqnarray*}
 where
 \begin{eqnarray*}
I_{0}^{(\chi_{b0})} & = & \int\limits _{-1}^{1}d\xi_{1}d\xi_{2}\frac{4+\xi_{1}^{2}+6\xi_{1}\xi_{2}+\xi_{2}^{2}}{4(1-\xi_{1}^{2})(1-\xi_{2}^{2})(1+\xi_{1}\xi_{2})^{2}}\phi_{1}^{L}(\xi_{1})\phi_{2}^{L}(\xi_{2}).
\end{eqnarray*}
The width of the $\chi_{b0}\to M_1 M_2$ decay is
\begin{eqnarray}
\Gamma(\chi_{b0} \to M_1 M_2) &=&
    \frac{2048\pi^3\alpha_s^4\beta}{2187} \frac{\langle O_P^{bb}\rangle}{m_b^4} \left[
        \frac{f_1^L f_2^L}{m_b^2}I_0^{(\chi_{b0})}
    \right]^2.
\end{eqnarray}
One should remember, that in the case of the identical final mesons the width
should be divided by  2.
Within NRQCD one can get the following expressions for the width
\begin{eqnarray*}
\Gamma(\chi_{b0}\to M_1 M_2) &=& \frac{1024\pi^3\alpha_s^4\beta}{2187}\frac{\left<O_P^{bb}\right>}{m_b^4}
  \left[\frac{f_1^\mathrm{NRQCD} f_2^\mathrm{NRQCD}}{m_b^2}\right]^2 F(\chi_{b0} \to M_1 M_2),
\end{eqnarray*}
where
\begin{eqnarray*}
F[\chi_{b0} \to \eta_c \chi_{c1}] &=& 4-16 r,\\
F[\chi_{b0} \to \chi_{c0} \chi_{c2}] &=& \frac{1}{9}-\frac{28 r}{9}+\frac{92 r^2}{3}-\frac{1120 r^3}{9}+\frac{1600 r^4}{9},\\
F[\chi_{b0} \to \eta_c \eta_c] &=& 1+4 r+4 r^2,\\
F[\chi_{b0} \to \psi \psi] &=& 1-4 r+12 r^2,\\
F[\chi_{b0} \to h_c h_c] &=& \frac{1}{4}-10 r^2-32 r^3+272 r^4,\\
F[\chi_{b0} \to \chi_{c0} \chi_{c0}] &=& \frac{1}{36}-\frac{16 r}{9}+28 r^2+\frac{128 r^3}{9}+\frac{16 r^4}{9},\\
F[\chi_{b0} \to \chi_{c1} \chi_{c1}] &=& 4-56 r+\frac{537 r^2}{2}-260 r^3+72 r^4,\\
F[\chi_{b0} \to \chi_{c2} \chi_{c2}] &=& \frac{1}{9}-\frac{28 r}{9}+\frac{179 r^2}{6}-\frac{340 r^3}{9}+\frac{1480 r^4}{9}
\end{eqnarray*}

{\bf The decays of the $\chi_{b1}$ meson.}
The decays $\chi_{b1} \to  h_{c}~ J/\psi (\psi'),~~
\eta_{c} (\eta_c')~ \chi_{c0}(\chi_{c2}),~~\chi_{c1}~\chi_{c0}(\chi_{c2})$ are the leading twist decays of
the $\chi_{b1}$ meson.
The amplitude for the processes equals
\begin{eqnarray*}
\mathcal{M}(\chi_{b1}\to M_{1}M_{2}) & = & \frac{64\sqrt{2}\pi^{2}\alpha_{s}^{2}}{27}\sqrt{\frac{\left\langle O^{b\bar b}_{1}\right\rangle _{P}}{m_{b}^{3}}}\frac{f_{1}^{L}f_{2}^{L}}{m_{b}^{2}}I_{0}^{(\chi_{b1})},
\end{eqnarray*}
where
\begin{eqnarray*}
I_{0}^{(\chi_{b1})} & = & \int\limits _{-1}^{1}d\xi_{1}d\xi_{2}\frac{\xi_{1}-\xi_{2}}{(1-\xi_{1}^{2})(1-\xi_{2}^{2})(1+\xi_{1}\xi_{2})}\phi_{1}^{L}(\xi_{1})\phi_{2}^{L}(\xi_{2}).
\end{eqnarray*}
The width of the $\chi_{b1}\to M_{1}M_{2}$ decay is
\begin{eqnarray*}
\Gamma(\chi_{b1}\to M_{1}M_{2}) & = & \frac{256\pi^{3}\alpha_{s}^{4}\beta}{2187}\frac{\left\langle O^{b\bar b}_{1}\right\rangle _{P}}{m_{b}^{4}}\left[\frac{f_{1}^{L}f_{2}^{L}}{m_{b}^{2}}I_{0}^{(\chi_{b1})}\right]^{2}.
\end{eqnarray*}

Within NRQCD one can get the following expressions for the width
\begin{eqnarray*}
\Gamma(\chi_{b1}\to M_1 M_2) &=& \frac{256\pi^3\alpha_s^4\beta}{2187}\frac{\left<O_P^{bb}\right>}{m_b^4}
  \left[\frac{f_1^\mathrm{NRQCD} f_2^\mathrm{NRQCD}}{m_b^2}\right]^2 F(\chi_{b1} \to M_1 M_2),
\end{eqnarray*}
where
\begin{eqnarray*}
F[\chi_{b1} \to h_c \psi] &=& 1+4 r-32 r^2,\\
F[\chi_{b1} \to \eta_c \chi_{c0}] &=& \frac{1}{3}-\frac{4 r}{3},\\
F[\chi_{b1} \to \eta_c \chi_{c2}] &=& \frac{2}{3}+\frac{4 r}{3}-16 r^2,\\
F[\chi_{b1} \to \chi_{c0} \chi_{c1}] &=& \frac{2}{3}+\frac{124 r}{3}+24 r^2,\\
F[\chi_{b1} \to \chi_{c2} \chi_{c1}] &=& \frac{4}{3}-\frac{16 r}{3}-8 r^2+192 r^3
\end{eqnarray*}

{\bf The decays of the $\chi_{b2}$ meson.}
The decays $\chi_{b2}  \to  \eta_{c} (\eta_{c}')~\chi_{c1}$, $\eta_{c}(\eta_{c}')~ \eta_{c}(\eta_{c}')$,
$ J/\psi(\psi') J/\psi(\psi'$), $h_{c}h_{c}$, $\chi_{c0}\chi_{c2}$, $\chi_{c0}~\chi_{c0}$,
$\chi_{c1}\chi_{c1}$, $\chi_{c2}~\chi_{c2}$, $h_{c}~J/\psi(\psi')$, $\chi_{c1}~\chi_{c2}$  are the leading twist decays of the $\chi_{b2}$ meson.
The decay of this meson is more complicated than the decays described above.
The point is that large spin of the $\chi_{c2}$ meson opens the possibility
to produce transversely polarized charmonia ($\lambda_1=-\lambda_2=\pm 1$)
at the leading twist approximation. So, contrary to the bottomonia decays
discussed above in some decays of the $\chi_{c2}$ meson there can be two
different polarizations of the final charmonia:  $\lambda_1=\lambda_2=0$ or
 $\lambda_1=-\lambda_2=\pm 1$. It should be noted here that it depends on
the process what possibilities are realized at the leading twist.
For instance, for the decay $\chi_{c2} \to \eta_c \chi_{c1}$ only $\lambda_1=\lambda_2=0$
is allowed, for the decay $\chi_{c2} \to J/\Psi J/\Psi$ both possibilities
$\lambda_1=\lambda_2=0, \lambda_1=-\lambda_2=\pm 1$ are allowed.
The state $\lambda_1=\lambda_2=0$ is forbidden for the decay $\chi_{b2} \to  h_{c}~J/\psi$
since in this case naturalness is not conserved. So, the only allowed possibility is
$\lambda_1=-\lambda_2=\pm 1$.

The amplitude for the decays of the $\chi_{b2}$ meson into pair of the longitudinally
polarized mesons ($\lambda_1=\lambda_2=0$) equals
\begin{eqnarray*}
\mathcal{M}(\chi_{b2}\to M_{1}^{L}M_{2}^{L}) & = & \frac{128\sqrt{2}\pi^{2}\alpha_{s}^{2}}{27\sqrt{3}}\sqrt{\frac{
\left\langle O^{b\bar b}_{1}\right\rangle _{P}}{m_{b}^{3}}}\frac{f_{1}^{L}f_{2}^{L}}{m_{b}^{2}}I_{0}^{(\chi_{b2})},
\end{eqnarray*}
where
\begin{eqnarray*}
I_{0}^{(\chi_{b2})} & = &
\int\limits_{-1}^{1}d\xi_{1}d\xi_{2}
\frac{2-\xi_{1}^{2}-\xi_{2}^{2}}{2\left(1-\xi_{1}^{2}\right)\left(1-\xi_{2}^{2}\right)\left(1+\xi_{1}\xi_{2}\right)^{2}}
\phi_{1}^{L}\left(\xi_{1}\right)\phi_{2}^{L}\left(\xi_{2}\right).
\end{eqnarray*}
If both final mesons can have nonzero helicity $\lambda_1=-\lambda_2=\pm 1$, one should take
into account the decays into transversely polarized particles. The amplitude of these decays is
\begin{eqnarray*}
\mathcal{M}(\chi_{b2}\to M_{1}^{T}M_{2}^{T}) & =- & \frac{256\pi^{2}\alpha_{s}^{2}}{27}\sqrt{\frac{\left\langle O^{b\bar b}_{1}\right\rangle _{P}}{m_{b}^{3}}}\frac{f_{1}^{T}f_{2}^{T}}{m_{b}^{2}}I_{2}^{(\chi_{b2})},
\end{eqnarray*}
 where
 \begin{eqnarray*}
I_{2}^{(\chi_{b2})} & = & \int\limits _{-1}^{1}d\xi_{1}d\xi_{2}\frac{1}{(1-\xi_{1}^{2})(1-\xi_{2}^{2})(1+\xi_{1}\xi_{2})}\phi_{1}^{T}(\xi_{1})\phi_{2}^{T}(\xi_{2}).
\end{eqnarray*}
The width of the $\chi_{b2}\to M_1 M_2$ decay is
\begin{eqnarray}
\Gamma(\chi_{b2}\to M_1 M_2) &=&
    \frac{1024\pi^3 \alpha_s^4\beta}{10935}\frac{\langle O_P^{bb}\rangle}{m_b^4}\left\{
        \left[\frac{f_1^L f_2^L}{m_b^2} I_0^{(\chi_{b2})}\right]^2+
        12 \left[\frac{f_1^T f_2^T}{m_b^2} I_2^{(\chi_{b2})}\right]^2
    \right\}
\end{eqnarray}

Within NRQCD one can get the following expression for the width of the process $\chi_{b2}\to M_1 M_2$
\begin{eqnarray*}
\Gamma(\chi_{b2}\to M_1 M_2) &=& \frac{512\pi^3\alpha_s^4\beta}{10935}\frac{\left<O_P^{bb}\right>}{m_b^4}
  \left[\frac{f_1^\mathrm{NRQCD} f_2^\mathrm{NRQCD}}{m_b^2}\right]^2 F(\chi_{b2} \to M_1 M_2)
\end{eqnarray*}
where
\begin{eqnarray*}
F[\chi_{b2} \to \eta_c \chi_{c1}] &=& 4+8 r-96 r^2,\\
F[\chi_{b2} \to \chi_{c0} \chi_{c2}] &=& \frac{16}{9}+\frac{872 r}{9}+\frac{208 r^2}{3}-\frac{544 r^3}{9}+\frac{4288 r^4}{9},\\
F[\chi_{b2} \to \eta_c \eta_c] &=& 1-8 r+16 r^2,\\
F[\chi_{b2} \to \psi \psi] &=& 13+56 r+48 r^2,\\
F[\chi_{b2} \to h_c h_c] &=& 16-132 r+488 r^2-944 r^3+800 r^4,\\
F[\chi_{b2} \to \chi_{c0} \chi_{c0}] &=& \frac{4}{9}-\frac{16 r}{9}-\frac{16 r^2}{3}+\frac{128 r^3}{9}+\frac{256 r^4}{9},\\
F[\chi_{b2} \to \chi_{c1} \chi_{c1}] &=& 7-44 r-30 r^2+340 r^3+264 r^4,\\
F[\chi_{b2} \to \chi_{c2} \chi_{c2}] &=& \frac{43}{9}+\frac{44 r}{9}-286 r^2+\frac{6212 r^3}{9}+\frac{5032 r^4}{9},\\
F[\chi_{b2} \to h_c \psi] &=& 24-72 r-96 r^2,\\
F[\chi_{b2} \to \chi_{c1} \chi_{c2}] &=& 6+4 r^2-488 r^3+1072 r^4
\end{eqnarray*}

\section{Numerical results.}

In order to obtain numerical results from the presented above analytical expressions
the following numerical parameters are needed.

In this paper we are going to use the models of the charmonia DAs proposed in papers
\cite{Braguta:2006wr, Braguta:2007fh, Braguta:2007tq, Braguta:2008qe}.
For the strong coupling constant we use one-loop expression
\begin{eqnarray*}
  \alpha_s(\mu) &=& \frac{4\pi}{b_0 \ln(\mu^2/\Lambda_\mathrm{QCD}^2)},
\end{eqnarray*}
where $b_0=25/3$ and $\Lambda_\mathrm{QCD}=0.2$ GeV.

In the calculation the following values of the constants $f^{L,T}_i$ defined in equations (\ref{constS}), (\ref{constP})
will be used
\begin{eqnarray}
f^L_{\eta_c} & = & 0.373 \pm 0.064 \,\mathrm{GeV}, \nonumber \\
f^L_{J/\psi} & = & 0.416 \pm 0.005\,\mathrm{GeV},\qquad f^T_{J/\psi}(M_{J/\Psi})  =  0.379 \pm 0.021  \,\mathrm{GeV}\nonumber \\
f^L_{\eta_c(2S)} &=& 0.261 \pm 0.077 \,\mathrm{GeV},\nonumber \\
f^L_{\psi(2S)} &=& 0.303 \pm 0.003 \,\mathrm{GeV},\,\qquad f^T_{\psi(2S)}(M_{J/\Psi}) = 0.261 \pm 0.042  \,\mathrm{GeV}, \nonumber \\
f^L_{\chi_{c0}}(M_{J/\Psi}) & = & 0.093 \pm 0.017\,\mathrm{GeV},\nonumber \\
f^L_{h_c}(M_{J/\Psi}) & = & 0.160 \pm 0.015  \,\mathrm{GeV},\qquad f^T_{h_c}(M_{J/\Psi})  =  0.179 \pm 0.032 \,\mathrm{GeV} \nonumber \\
f^L_{\chi_{c1}} & = & 0.272 \pm 0.048 \,\mathrm{GeV},\qquad f^T_{\chi_{c1}}(M_{J/\Psi})  =  0.111 \pm 0.020 \,\mathrm{GeV}, \nonumber \\
f^L_{\chi_{c2}}(M_{J/\Psi}) & = & 0.131 \pm 0.023 \,\mathrm{GeV},\qquad f^T_{\chi_{c2}}(M_{J/\Psi})  =  0.157 \pm 0.028 \,\mathrm{GeV}.
\label{const_values}
\end{eqnarray}
The values of the constants $f^L_{J/\psi}, f^L_{\psi(2S)}$ can be extracted from the leptonic decay widths
of the $J/\psi$ and $\psi(2S)$ mesons. The values of the constants $f^L_{\eta_c}, f^L_{\eta_c(2S)}$ were
calculated in paper \cite{Braguta:2008tg}. The values of the constants $f^T_{J/\psi}, f^T_{\psi(2S)}$ can be found
in paper \cite{Braguta:2007ge}. The values of the constants of the $P$-wave charmonia mesons
can be found in paper \cite{Braguta:2008qe}. It should be noted that the constants
$f^T_{J/\psi, \psi(2S)}, f^L_{\chi_{c0}}, f^{L,T}_{h_c}, f^T_{\chi_{c1}}, f^{L,T}_{\chi_{c2}}$
depend on the renormalization scale. As it is seen from formulas (\ref{const_values})
these constants are defined at the scale $\mu=M_{J/\Psi}$. The anomalous dimensions of these constants,
which govern the evolution, can be found in papers \cite{Braguta:2007ge, Braguta:2008qe}.

The values of NRQCD matrix elements $\left<O_1^{b\bar b}\right>_S, \left<O_1^{b\bar b}\right>_P$
can be expressed through the bottomonia radial wave function $R(r)$ as follows
\begin{eqnarray}
\left<O_1^{b\bar b}\right>_S = \frac {3} {2 \pi} |R_S(0)|^2,~~~~~~
\left<O_1^{b\bar b}\right>_P = \frac {9} {2 \pi} |R_P(0)|^2.
\end{eqnarray}
In this paper the values of the $|R_S(0)|^2, |R_P(0)|^2$ will be determined from
the Buchmuller-Tye \cite{Buchmuller:1980su, Eichten:1995ch} potential model. Thus one gets
\begin{eqnarray*}
\left<O_1^{b\bar b}\right>_S &=& 3.1\,\mathrm{GeV}^3,\qquad \left<O_1^{b\bar b}\right>_P = 2.0\,\mathrm{GeV}^5.
\end{eqnarray*}
In the forthcoming analysis we are not going to take into the account the
uncertainties in the values of the NRQCD matrix elements for bottomonia mesons
since these uncertainties are not very important.

In the calculation we also need the values of  constants (\ref{fsnr}), (\ref{fpnr}) for the
charmonia mesons. At the leading order approximation in $\alpha_s$ and relative velocity
these constants can be determined from the leptonic decay widths
\begin{eqnarray}
\Gamma(V \to e^+e^-)=\frac {4 q_c^2 \pi \alpha^2} {3 M_V} \bigl [ f_S^\mathrm{NRQCD} \bigr ]^2.
\label{vee}
\end{eqnarray}
Using experimental results $\Gamma( J/\Psi \to e^+ e^-)=5.55 ~ \mbox{KeV},
\Gamma( \psi(2S) \to e^+ e^-)=2.48~ \mbox{KeV}$ \cite{Yao:2006px} one gets
\begin{eqnarray}
\bigl [ f_{1S}^\mathrm{NRQCD} \bigr ]^2 = 0.17 \pm 0.06  \,\mathrm{GeV}^2,~~~
\bigl [ f_{2S}^\mathrm{NRQCD} \bigr ]^2 = 0.09 \pm 0.05 \,\mathrm{GeV}^2.
\label{nrS}
\end{eqnarray}
To determine the constant $f_{P}^\mathrm{NRQCD}$ one can use the decay width $\chi_{c0} \to \gamma \gamma$
\begin{eqnarray}
\Gamma(\chi_{c0} \to \gamma \gamma)=\frac {12 q_c^4 \pi \alpha^2} {M_{\chi_{c0}}} \bigl [ f_P^\mathrm{NRQCD} \bigr ]^2.
\label{cgaga}
\end{eqnarray}
Using experimental results $Br( \chi_{c0} \to \gamma \gamma)=2.35 \times 10^{-4}$ \cite{Yao:2006px} one gets
\begin{eqnarray}
\bigl [ f_{P}^\mathrm{NRQCD} \bigr ]^2 =  0.021 \pm 0.008 \,\mathrm{GeV}^2.
\label{nrP}
\end{eqnarray}
The uncertainties in  values (\ref{nrS}), (\ref{nrP}) were calculated as follows.
In NRQCD there are relativistic and radiative corrections to formulas
(\ref{vee}) and (\ref{cgaga}). The relativistic corrections can be estimated as
$\langle v^2 \rangle_{1S} =0.21$ \cite{Braguta:2006wr},
$\langle v^2 \rangle_{2S} =0.54$ \cite{Braguta:2007tq}, $\langle v^2 \rangle_{P} =0.3$ \cite{Braguta:2008qe}.
The radiative corrections can be estimated as $\sim \alpha_s(M_{J/\Psi})=0.25$.
Adding these uncertainties in quadrature one estimates the errors of the calculation.

The last parameter needed for calculation of the bottomonia decay widths is
the pole masses of $c$-quarks. For the $c$-quark we take $m_c=1.4 \pm 0.2$ GeV.
As was noted before for the mass of $b$-quark we take the value $m_b=M/2$, where $M$ is the mass of the decaying bottominium.

In Table \ref{tab:Widths} we present numerical results for the widths of
exclusive bottomonia decays into pair of charmonia mesons. In the second
and third  columns of this table the results of NRQCD and light cone
formalism are shown. In the fourth column we present the branching fractions of the
considered decays. To estimate these fractions we use the following expressions
for the total widths of bottomonia mesons \cite{Bodwin:2007zf}:
\begin{eqnarray}
\Gamma_{\eta_b} &=& \frac{4\pi\alpha_s^2}{9}\frac{\langle O_S^{bb}\rangle}{m_b^2}\approx 9.9\ \mathrm{MeV},\\
\Gamma_{\chi_{b0}} & = &
    \frac{3C_F}{N_c}\pi\alpha_s^2 \frac{\langle O_P^{bb}\rangle}{m_b^4} + \frac{n_f}{3}\pi\alpha_s^2 \frac{\langle O_8\rangle}{m_b^2}
    \approx 0.80\ \mathrm{MeV},\\
\Gamma_{\chi_{b2}} & = &
    \frac{4C_F}{5N_c}\pi\alpha_s^2 \frac{\langle O_P^{bb}\rangle}{m_b^4} + \frac{n_f}{3}\pi\alpha_s^2 \frac{\langle O_8\rangle}{m_b^2}
    \approx 0.2\ \mathrm{MeV},\\
\Gamma{\chi_{b1}} &=& \frac{C_F\alpha_s^3}{N_c}\left[
        \left(\frac{587}{54}-\frac{317}{288}\pi^2\right)C_A+
        \left(-\frac{16}{27}-\frac{4}{9}\ln\frac{\Lambda}{2m_b}\right)n_f
\right]\frac{\langle O_P^{bb}\rangle}{m_b^4}
    +\frac{n_f}{3}\pi\alpha_s^2 \frac{\langle O_8\rangle}{m_b^2} \approx\nonumber \\
&\approx& 0.13\ \mathrm{MeV},
\end{eqnarray}
where $n_f=4$ is number of active flavors, $\Lambda=200$ MeV, and
$\langle O_8\rangle\approx 0.0021 \langle O_P^{bb}\rangle$ is the color octet matrix element for  the $P$-wave bottomonia mesons.
Since experimentally charmonia mesons are observed in their decays into $J/\psi$ meson, it
is interesting to know the widths of such processes. It is clear that they are equal to
\begin{eqnarray*}
\Gamma[ (b\bar b) \to (c\bar c)_1 (c\bar c)_2 \to J/\psi J/\psi + X] &=&
   \Gamma[ (b\bar b) \to (c\bar c)_1 (c\bar c)_2] \mathrm{Br}[(c\bar c)_1 \to J/\psi+X]
   \mathrm{Br}[(c\bar c)_2 \to J/\psi+X]
\end{eqnarray*}
These values are shown in the last column of Table \ref{tab:Widths}.

\begin{table}
\begin{center}
\begin{tabular}{|c|c|c|c|c|}
\hline
reaction & $\Gamma_{\mathrm{NRQCD}}$, eV &$\Gamma_{\mathrm{LC}}$, eV &  $\mathrm{Br}_{\mathrm{LC}}, 10^{-5}$  &$\mathrm{Br}_{\mathrm{LC}}(\psi\psi), 10^{-5}$ \\ 
\hline
 $\eta_b \to h_c \psi$ &  $16.^{+2.3}_{-1.5} \pm 8.4 \pm 8.1$  &  $32. \pm 2.6 \pm 6.1 \pm 8.2$  &  $ 0.33$ &  --- \\ 
 $\eta_b \to h_c \psi(2S)$ &  $7.8^{+1.1}_{-0.72} \pm 6.5 \pm 3.9$  &  $16. \pm 1.4 \pm 3.1 \pm 4.2$  &  $ 0.17$ &  --- \\ 
 $\eta_b \to \eta_c \chi_{c0}$ &  $13.^{+3.5}_{-2.7} \pm 6.8 \pm 6.5$  &  $9.1 \pm 0.73 \pm 4.6 \pm 2.3$  &  $ 0.092$ &  --- \\ 
 $\eta_b \to \eta_c(2S) \chi_{c0}$ &  $6.3^{+1.7}_{-1.3} \pm 5.2 \pm 3.1$  &  $4.3 \pm 0.36 \pm 3. \pm 1.1$  &  $ 0.043$ &  --- \\ 
 $\eta_b \to \eta_c \chi_{c2}$ &  $3.6^{+1.1}_{-1.1} \pm 8.4 \pm 1.8$  &  $18. \pm 1.4 \pm 8.7 \pm 4.5$  &  $ 0.18$ &  --- \\ 
 $\eta_b \to \eta_c(2S) \chi_{c2}$ &  $1.7^{+0.54}_{-0.53} \pm 4.1 \pm 0.86$  &  $8.2 \pm 0.7 \pm 5.6 \pm 2.1$  &  $ 0.083$ &  --- \\ 
 $\eta_b \to \chi_{c0} \chi_{c1}$ &  $2.3^{+0.21}_{-0.29} \pm 2.2 \pm 1.2$  &  $4.4 \pm 0.38 \pm 2.3 \pm 1.1$  &  $ 0.045$ &  $ 2.1\times 10^{-4}$\\ 
 $\eta_b \to \chi_{c1} \chi_{c2}$ &  $0.93^{+0.22}_{-0.21} \pm 2.9 \pm 0.46$  &  $8.6 \pm 0.73 \pm 4.3 \pm 2.2$  &  $ 0.087$ &  $ 0.0062$\\ 
\hline
 $\chi_{b0} \to \eta_c \chi_{c1}$ &  $1.9^{+0.23}_{-0.27} \pm 1.9 \pm 0.93$  &  $9.8 \pm 0.25 \pm 4.8 \pm 2.5$  &  $ 1.2$ &  --- \\ 
 $\chi_{b0} \to \eta_c(2S) \chi_{c1}$ &  $0.9^{+0.11}_{-0.13} \pm 1.1 \pm 0.45$  &  $5.9 \pm 1. \pm 4. \pm 1.5$  &  $ 0.73$ &  --- \\ 
 $\chi_{b0} \to \chi_{c0} \chi_{c2}$ &  $0.00015^{+0.0007}_{-0.00014} \pm 0.038 \pm 7.6\times 10^{-5}$  &  $0.14 \pm 0.034 \pm 0.07 \pm 0.034$  &  $ 0.017$ &  $ 4.5\times 10^{-5}$\\ 
 $\chi_{b0} \to \eta_c \eta_c$ &  $7.9^{+0.69}_{-0.57} \pm 5.6 \pm 4.$  &  $10. \pm 0.45 \pm 4.9 \pm 2.5$  &  $ 1.3$ &  --- \\ 
 $\chi_{b0} \to \eta_c \eta_c(2S)$ &  $7.8^{+0.68}_{-0.56} \pm 7.5 \pm 3.9$  &  $12. \pm 2.1 \pm 8.3 \pm 3.$  &  $ 1.5$ &  --- \\ 
 $\chi_{b0} \to \eta_c(2S) \eta_c(2S)$ &  $1.9^{+0.16}_{-0.14} \pm 2.8 \pm 0.94$  &  $3.6 \pm 1.4 \pm 3. \pm 0.91$  &  $ 0.45$ &  --- \\ 
 $\chi_{b0} \to \psi \psi$ &  $4.3^{+0.28}_{-0.25} \pm 5.7 \pm 2.2$  &  $15. \pm 0.68 \pm 0.51 \pm 3.8$  &  $ 1.9$ &  $ 1.9$\\ 
 $\chi_{b0} \to \psi \psi(2S)$ &  $4.3^{+0.28}_{-0.25} \pm 6.3 \pm 2.1$  &  $20. \pm 3.5 \pm 0.62 \pm 5.$  &  $ 2.5$ &  $ 1.4$\\ 
 $\chi_{b0} \to \psi(2S) \psi(2S)$ &  $1.^{+0.068}_{-0.06} \pm 1.9 \pm 0.52$  &  $6.5 \pm 2.5 \pm 0.18 \pm 1.6$  &  $ 0.81$ &  $ 0.27$\\ 
 $\chi_{b0} \to h_c h_c$ &  $0.014^{+0.0025}_{-0.0035} \pm 0.021 \pm 0.0071$  &  $0.3 \pm 0.074 \pm 0.079 \pm 0.075$  &  $ 0.037$ &  --- \\ 
 $\chi_{b0} \to \chi_{c0} \chi_{c0}$ &  $0.006^{+0.0076}_{-0.0041} \pm 0.022 \pm 0.003$  &  $0.035 \pm 0.0087 \pm 0.018 \pm 0.0088$  &  $ 0.0044$ &  $ 7.4\times 10^{-7}$\\ 
 $\chi_{b0} \to \chi_{c1} \chi_{c1}$ &  $0.087^{+0.037}_{-0.025} \pm 0.63 \pm 0.043$  &  $2.4 \pm 0.12 \pm 1.2 \pm 0.6$  &  $ 0.3$ &  $ 0.038$\\ 
 $\chi_{b0} \to \chi_{c2} \chi_{c2}$ &  $0.0032^{+0.0038}_{-0.0012} \pm 0.035 \pm 0.0016$  &  $0.13 \pm 0.033 \pm 0.066 \pm 0.033$  &  $ 0.017$ &  $ 6.8\times 10^{-4}$\\ 
 \hline
$\chi_{b1} \to h_c \psi$ &  $0.18^{+0.0016}_{-0.0077} \pm 0.13 \pm 0.091$  &  $0.88 \pm 0.078 \pm 0.17 \pm 0.22$  &  $ 0.68$ &  --- \\ 
 $\chi_{b1} \to h_c \psi(2S)$ &  $0.089^{+0.00076}_{-0.0037} \pm 0.086 \pm 0.045$  &  $0.67 \pm 0.18 \pm 0.13 \pm 0.17$  &  $ 0.52$ &  --- \\ 
 $\chi_{b1} \to \eta_c \chi_{c0}$ &  $0.038^{+0.0048}_{-0.0055} \pm 0.038 \pm 0.019$  &  $0.25 \pm 0.022 \pm 0.12 \pm 0.061$  &  $ 0.19$ &  --- \\ 
 $\chi_{b1} \to \eta_c(2S) \chi_{c0}$ &  $0.019^{+0.0023}_{-0.0027} \pm 0.022 \pm 0.0093$  &  $0.17 \pm 0.046 \pm 0.12 \pm 0.043$  &  $ 0.13$ &  --- \\ 
 $\chi_{b1} \to \eta_c \chi_{c2}$ &  $0.11^{+0.0031}_{-0.0066} \pm 0.075 \pm 0.055$  &  $0.48 \pm 0.042 \pm 0.24 \pm 0.12$  &  $ 0.37$ &  --- \\ 
 $\chi_{b1} \to \eta_c(2S) \chi_{c2}$ &  $0.054^{+0.0015}_{-0.0032} \pm 0.051 \pm 0.027$  &  $0.33 \pm 0.089 \pm 0.23 \pm 0.083$  &  $ 0.26$ &  --- \\ 
 $\chi_{b1} \to \chi_{c0} \chi_{c1}$ &  $0.08^{+0.022}_{-0.018} \pm 0.061 \pm 0.04$  &  $0.12 \pm 0.015 \pm 0.06 \pm 0.03$  &  $ 0.091$ &  $ 4.2\times 10^{-4}$\\ 
 $\chi_{b1} \to \chi_{c1} \chi_{c2}$ &  $0.018^{+0.0015}_{-0.00087} \pm 0.028 \pm 0.0091$  &  $0.23 \pm 0.03 \pm 0.11 \pm 0.057$  &  $ 0.18$ &  $ 0.013$\\ 
 \hline
 $\chi_{b2} \to \eta_c \chi_{c1}$ &  $0.26^{+0.0073}_{-0.015} \pm 0.18 \pm 0.13$  &  $0.63 \pm 0.011 \pm 0.31 \pm 0.16$  &  $ 0.31$ &  --- \\ 
 $\chi_{b2} \to \eta_c(2S) \chi_{c1}$ &  $0.13^{+0.0036}_{-0.0075} \pm 0.12 \pm 0.064$  &  $0.35 \pm 0.044 \pm 0.24 \pm 0.086$  &  $ 0.17$ &  --- \\ 
 $\chi_{b2} \to \chi_{c0} \chi_{c2}$ &  $0.076^{+0.02}_{-0.017} \pm 0.058 \pm 0.038$  &  $0.049 \pm 0.0075 \pm 0.025 \pm 0.012$  &  $ 0.025$ &  $ 6.5\times 10^{-5}$\\ 
 $\chi_{b2} \to \eta_c \eta_c$ &  $0.26^{+0.069}_{-0.069} \pm 0.69 \pm 0.13$  &  $0.64 \pm 0.02 \pm 0.31 \pm 0.16$  &  $ 0.32$ &  --- \\ 
 $\chi_{b2} \to \eta_c(2S) \eta_c$ &  $0.26^{+0.068}_{-0.068} \pm 0.7 \pm 0.13$  &  $0.71 \pm 0.092 \pm 0.48 \pm 0.18$  &  $ 0.36$ &  --- \\ 
 $\chi_{b2} \to \eta_c(2S) \eta_c(2S)$ &  $0.062^{+0.016}_{-0.017} \pm 0.18 \pm 0.031$  &  $0.2 \pm 0.068 \pm 0.17 \pm 0.051$  &  $ 0.1$ &  --- \\ 
 $\chi_{b2} \to \psi \psi$ &  $9.7^{+0.87}_{-0.73} \pm 6.9 \pm 4.9$  &  $9.6 \pm 0.42 \pm 0.33 \pm 2.4$  &  $ 4.8$ &  $ 4.8$\\ 
 $\chi_{b2} \to \psi(2S) \psi$ &  $9.6^{+0.86}_{-0.72} \pm 9.3 \pm 4.8$  &  $11. \pm 1.9 \pm 0.35 \pm 2.8$  &  $ 5.7$ &  $ 3.3$\\ 
 $\chi_{b2} \to \psi(2S) \psi(2S)$ &  $2.3^{+0.21}_{-0.17} \pm 3.5 \pm 1.2$  &  $3.4 \pm 1.4 \pm 0.094 \pm 0.84$  &  $ 1.7$ &  $ 0.56$\\ 
 $\chi_{b2} \to h_c h_c$ &  $0.061^{+0.012}_{-0.012} \pm 0.17 \pm 0.031$  &  $0.48 \pm 0.034 \pm 0.13 \pm 0.12$  &  $ 0.24$ &  --- \\ 
 $\chi_{b2} \to \chi_{c0} \chi_{c0}$ &  $0.0021^{+0.00037}_{-0.00044} \pm 0.0037 \pm 0.0011$  &  $0.013 \pm 0.0019 \pm 0.0065 \pm 0.0032$  &  $ 0.0063$ &  $ 1.1\times 10^{-6}$\\ 
 $\chi_{b2} \to \chi_{c1} \chi_{c1}$ &  $0.026^{+0.0069}_{-0.0074} \pm 0.063 \pm 0.013$  &  $0.28 \pm 0.03 \pm 0.14 \pm 0.069$  &  $ 0.14$ &  $ 0.018$\\ 
 $\chi_{b2} \to \chi_{c2} \chi_{c2}$ &  $0.028^{+0.0038}_{-0.0052} \pm 0.042 \pm 0.014$  &  $0.54 \pm 0.11 \pm 0.27 \pm 0.13$  &  $ 0.27$ &  $ 0.011$\\ 
 $\chi_{b2} \to h_c \psi$ &  $1.1^{+0.12}_{-0.14} \pm 1. \pm 0.57$  &  $3.6 \pm 0.09 \pm 0.68 \pm 0.9$  &  $ 1.8$ &  --- \\ 
 $\chi_{b2} \to h_c \psi(2S)$ &  $0.56^{+0.057}_{-0.069} \pm 0.62 \pm 0.28$  &  $2.1 \pm 0.36 \pm 0.39 \pm 0.52$  &  $ 1.$ &  --- \\ 
 $\chi_{b2} \to \chi_{c1} \chi_{c2}$ &  $0.044^{+0.0008}_{-0.0015} \pm 0.036 \pm 0.022$  &  $0.49 \pm 0.1 \pm 0.24 \pm 0.12$  &  $ 0.25$ &  $ 0.018$\\ 
\hline
\label{tab:Widths}
\end{tabular}
\end{center}
\caption{The widths and branching fractions of the exclusive bottomonia decays
into pair of charmonium mesons. In the second column the NRQCD predictions are presented.
In the third and fourth the widths and branching fractions of the exclusive bottomonia
decays in LC formalism are  shown. The last column contains the branching
fractions of inclusive $J/\psi$-pair production through intermediate charmonium
states.
The symbol "---" in this column means, that this decay is forbidden 
(for example, decay $\eta_b\to\eta_c \chi_{c0}\to J/\psi J/\psi+X$ is absent since 
$\eta_c$ meson cannot decay into $J/\psi$) or its branching fraction is unknown (for example, in the case $\eta_b \to h_c J/\psi$)
}
\end{table}

Now let us discuss the uncertainties of the calculation. Before
we discuss how the uncertainties of the NRQCD prediction can be
estimated, one should recall the experience gained from double
charmonium production at B-factories. In this case the leading
order NRQCD predictions \cite{Braaten:2002fi, Liu:1, Liu:2} are approximately  by an order of magnitude
less than experimental results \cite{Abe:2004ww, Aubert:2005tj}.
Note also that measured values of the cross sections are much
larger than the leading order NRQCD predictions even if one
takes into account the possible uncertainties of the
approach \cite{Braaten:2002fi}. From this fact one can conclude
that it is rather difficult to calculate the uncertainties of NRQCD. So, the
uncertainties calculated in this paper can be considered only as
a very rough estimation of the real uncertainties.

\begin{figure}
\begin{centering}
\includegraphics[scale=0.6]{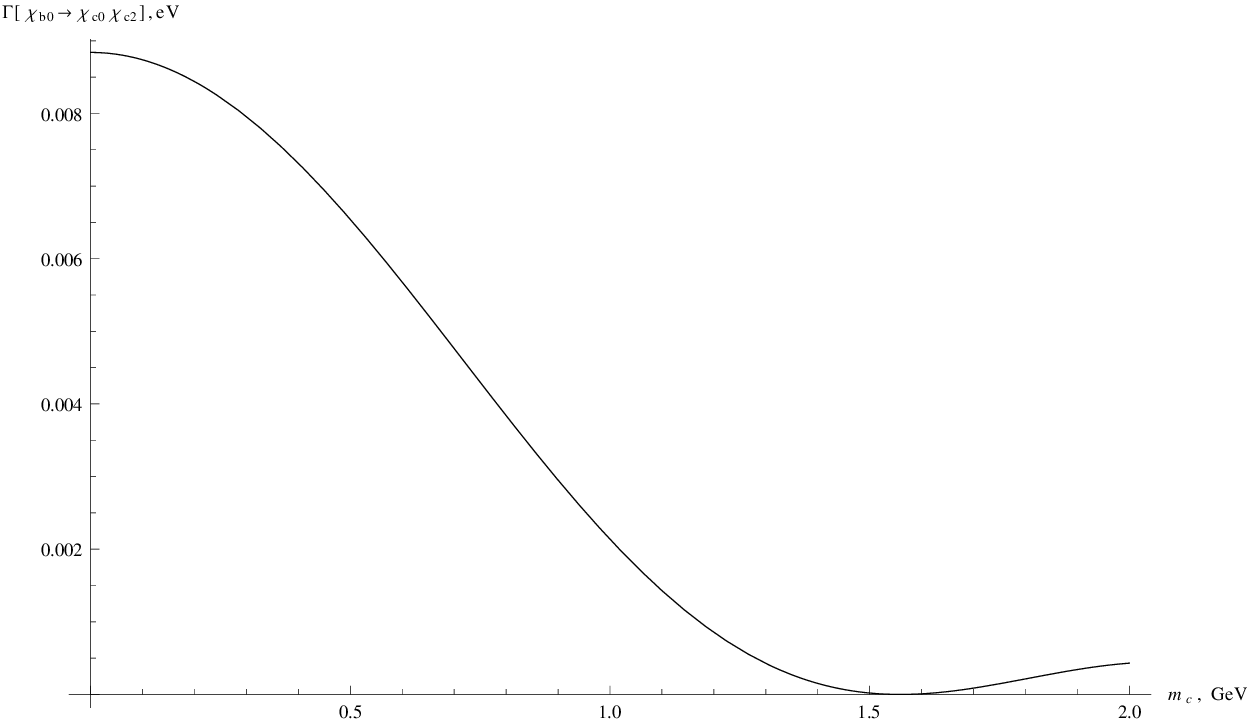}
\par\end{centering}
\caption{The width of the $\chi_{b0} \to \chi_{c0} \chi_{c2}$ decay as a function of
the mass of $c$-quark $m_c$.}
\end{figure}

Note also that in some decays ( see section III), the amplitudes contain
polynomials in $r$ with alternating signs of the coefficients
of these polynomials. One can expect that the uncertainty
of the calculation of such decays can be very large.
The decay $\chi_{b0} \to \chi_{c0} \chi_{c2}$ can be
considered as a dramatic demonstration of this point.
The width of this decay as a function of the mass of $c$-quark
is shown in Fig.2. It is seen from this figure that
the width $\chi_{b0} \to \chi_{c0} \chi_{c2}$ has zero
and minimum at $m_c \simeq 1.56$ GeV which is very near to the
pole mass of $c$-quark. Evidently, the uncertainty
in the width of this decay can be very large.

As the uncertainty of the leading order NRQCD prediction we take the uncertainty
which originates from the following sources:  uncertainty in the
pole mass of $c$  quark (the first error in the
second column of Tab. \ref{tab:Widths}),
uncertainty due to the values of constants (\ref{nrS}), (\ref{nrP})
(the second error in Tab. \ref{tab:Widths}),
uncertainty due to the unknown
 radiative corrections, which can be estimated as
$\sim \alpha_s(M_{b \bar b}) \log{m_b^2/m_c^2} \sim 50 \%$ (the third error in Tab. \ref{tab:Widths}).
The last uncertainty (the forth error in Tab. \ref{tab:Widths}) originates from the uncertainty
in the value of the $\Lambda=200 \pm 40$ MeV what corresponds to 
$\alpha_s(m_{\tau})=0.34 \pm 0.03$\cite{Yao:2006px}. 

The uncertainties of the results obtained within LC can be divided into the following groups:

{\bf 1.} The uncertainty in the models of the distribution amplitudes,
which can be estimated through the variation of the parameters of these models
(see papers \cite{Braguta:2006wr, Braguta:2007fh, Braguta:2007tq, Braguta:2008qe})
(the first error in the third column of Tab. \ref{tab:Widths}).

{\bf 2.} The uncertainty in the values of constants (\ref{const_values})
(the second error in the third column of Tab. \ref{tab:Widths})

{\bf 3.} The uncertainty due to the power corrections. This source of
uncertainty is very important and for many processes this is the main source of the uncertainty.
To estimate this source of the uncertainty we expand of the leading order NRQCD results
in the ratio $r=m_c^2/m_b^2$. The first term in this expansion
is the leading twist contribution which is reproduced within LC.
We take the next nonvanishing term in the $r$ expansion as the estimation
of the size of power corrections. (the third error in Tab. \ref{tab:Widths})

Applying this approach for the estimation of power corrections and looking to
the leading order NRQCD results (section III) one can separate all
processes into three groups. The first group contains the processes
(for instance, the decays $\eta_b \to h_c \psi, \chi_{b0} \to \eta_c \chi_{c1}$, )
for which power corrections, most probably, will not change the LC predictions dramatically.
One can expect that for this group of the decays LC predictions are reliable.
The second group contain the processes for which power corrections
are of order of $\sim 100 \%$ (for instance,
$\eta_b \to \eta_c \chi_{c0}, \chi_{b0} \to \chi_{c1} \chi_{c1}$ ).
For such processes our results are valid up to the factor of $\sim 2$.
The last group of processes are the processes for which power corrections are
large (for instance, $\chi_{b0} \to \chi_{c0} \chi_{c0}, \chi_{b1} \to \chi_{c0} \chi_{c1}$).
For these processes we can guess only the order of magnitude of the widths within LC.

{\bf 4.} The uncertainty due to the radiative corrections.
The main part of the radiative corrections to the amplitude -- the
leading logarithmic radiative corrections have been resummed within LC.  This fact allows us to estimate
the rest of the radiative corrections as  $\sim \alpha_s( E ) \sim 20 \%$.
This is very small uncertainty, so we don't show it in Tab. \ref{tab:Widths}.

{\bf 5.} The uncertainty due to the variation of $\Lambda=200 \pm 40$ MeV. 
(the forth  error in Tab. \ref{tab:Widths}). It should be noted also that 
in the calculation we took the scale of factorization $\mu=M_{b \bar b}/2$. 
However, one can take any scale $\mu \sim M_{b \bar b}$. The dependence of the
branching ratios on the $\mu$ are rather strong since $Br \sim \alpha_s^2$. 
For instance, if we change $\mu$ from $M_{b \bar b}/2$ to $M_{b \bar b}$ the
branching ratios will be changed by $40 -50$ \%.

It can be easily seen (see Table \ref{tab:comp}), that our predictions for branching fractions $\chi_{b0,2}\to2J/\psi$, $\chi_{b0,2}\to\psi\psi(2S)$ and $\chi_{b0,2}\to2\psi(2S)$ satisfy the upper bounds, that were set recently by Belle Collaboration \cite{Shen:2012ei}. Our results are also in reasonable agreement with numbers cited in \cite{Zhang:2011ng}.

\begin{table}
\begin{tabular}{|c|c|c|c|}
\hline
reaction & our results & \cite{Shen:2012ei} & \cite{Zhang:2011ng} \\
\hline
$\chi_{b0}\to2J/\psi$ &  $1.9\pm0.1\pm0.1\pm0.5$ &  $<7.1$ & $0.5$ \\
$\chi_{b2}\to2J/\psi$ &  $4.8\pm0.2\pm0.2\pm1.2$ &  $<4.5$ & $3.4$ \\
\hline
$\chi_{b0}\to J/\psi\ \psi(2S)$ &  $2.5\pm 0.4\pm0.1 \pm 0.6$ &  $<12$ & --- \\
$\chi_{b2}\to J/\psi\ \psi(2S)$ &  $5.7\pm 1.0\pm0.2 \pm 1.4$ &  $<4.9$ & --- \\
\hline
$\chi_{b0}\to 2\psi(2S)$ &  $0.81\pm 0.32\pm 0.02 \pm 0.2$ &  $<3.1$ & --- \\
$\chi_{b2}\to 2\psi(2S)$ &  $1.7 \pm 0.7\pm0.05 \pm 0.4$ &  $<1.6$ & --- \\
\hline
\end{tabular}
\caption{Branching fractions ($10^{-5}$) of scalar and tensor bottomonia decays into pair of vector charmonia}
\label{tab:comp}
\end{table}

\section{Conclusion.}

In this paper the leading twist double charmonium production in exclusive
bottomonia decays was considered. The decays of the C-odd bottomonia
are suppressed by the factor $\sim \alpha_s/\pi$ in comparison
to the decays of the C-even bottomonia. For this reason we
considered only the leading twist decays of the $C$-even bottomonia.
Applying light cone formalism with the models of the leading twist
charmonia distribution amplitudes \cite{Braguta:2006wr, Braguta:2007fh,
Braguta:2007tq, Braguta:2008qe} we calculated the amplitudes and the
widths of the corresponding processes. In addition, we calculated the
widths within the leading order NRQCD.

During the calculation we found that the production of the
longitudinally polarized $\chi_{c1}$ meson and transversely
polarized $h_c$ meson with $L_z \neq 0$ is nonzero already
at the leading twist approximation. This fact tells
us that the second selection rule ( see section II ),
which predicts the suppression of such processes, is violated.
We considered the mechanism which lies behind this
violation and found that this violation results
from the rather special Lorentz structure
of the corresponding distribution amplitudes.

\begin{acknowledgments}
This work was partially supported by the Russian
Foundation of Basic Research under Grant No. 07-02-
00417. The work of V. B. was partially supported by
CRDF Grant No. Y3-P-11-05 and president Grant
No. MK-140.2009.2. The work of A. V. L. was partially
supported by president Grant No. MK-110.2008.2, and
grants of the noncommercial foundation "Dynasty" and
the Russian Science Support Foundation.
\end{acknowledgments}

\newpage

\appendix

\section{Distribution amplitudes.}

The leading twist distribution amplitudes needed in the calculation can be defined as
follows:

{\bf for the pseudoscalar mesons $P=\eta_c, \eta_c'$:}
\begin{eqnarray*}
\left\langle P(p)\left|
\bar{Q}^i_{\alpha}(z) [z,-z]Q^j_{\beta}(-z)\right|0\right\rangle  & = &
(\hat p \gamma_5)_{\beta \alpha} \frac {f_P}{4} \frac{\delta_{ij}}{3}
\int\limits _{-1}^{1}d\xi e^{i\xi(pz)}\phi_{P}(\xi;\mu),
\end{eqnarray*}

{\bf for the vector mesons $V=J/\Psi, \psi'$:}
\begin{eqnarray*}
\left\langle V(p,\epsilon_{\lambda=0}) \left| \bar{Q}^i_{\alpha} (z) [z,-z] Q^j_{\beta}(-z) \right|0\right\rangle  & = &
(\hat p )_{\beta \alpha}  \frac {f_{V}^{L}} {4} \frac{\delta_{ij}}{3}
 \int\limits _{-1}^{1}d\xi e^{i\xi(pz)}\phi_{V}^{L}(\xi;\mu)
 \\
\left\langle V(p,\epsilon_{\lambda=\pm 1}) \left| \bar{Q}^i_{\alpha}(z) [z,-z] Q^j_{\beta}(-z) \right|0\right\rangle   & = &
(\hat p \hat {\epsilon} )_{\beta \alpha}  \frac {f_{V}^{T}} {4}  \frac{\delta_{i j}}{3}
  \int\limits _{-1}^{1}d\xi e^{i\xi(pz)}\phi_{V}^{T}(\xi;\mu)
\end{eqnarray*}

{\bf  for the $\chi_{c0}$-meson:}
\begin{eqnarray*}
\left\langle \chi_{c0}(p)\left|\bar{Q}^i_{\alpha} (z) [z,-z] Q^j_{\beta}(-z)\right|0\right\rangle  & = &
(\hat p)_{\beta \alpha}  \frac{f_{\chi0}^{L}}{4}  \frac{\delta_{ij}}{3}
\int\limits _{-1}^{1}d\xi e^{i\xi(pz)}\phi_{\chi0}^{L}(\xi;\mu)
\end{eqnarray*}

{\bf  for the $\chi_{c1}$-meson:}
\begin{eqnarray*}
\left\langle \chi_{c1}(p,\epsilon_{\lambda=0})\left|\bar{Q}^i_{\alpha} (z) [z,-z] Q^j_{\beta}(-z)\right|0\right\rangle  & = &
(\hat p \gamma_5)_{\beta \alpha} \frac{f_{\chi1}^{L}}{4}  \frac{\delta_{ij}}{3}
\int\limits _{-1}^{1}d\xi e^{i\xi(pz)}\phi_{\chi1}^{L}(\xi;\mu),
\\
\left\langle \chi_{c1}(p,\epsilon_{\lambda=\pm1})\left|\bar{Q}^i_{\alpha} (z) [z,-z] Q^j_{\beta}(-z)\right|0\right\rangle  & = &
(\hat p \hat \epsilon \gamma_5)_{\beta\alpha} \frac{f_{\chi1}^{T}}{4} \frac{\delta_{ij}}{3}
\int\limits _{-1}^{1}d\xi e^{i\xi(pz)}\phi_{\chi1}^{T}(\xi;\mu)\end{eqnarray*}

{\bf  for the $h_{c}$-meson:}
\begin{eqnarray*}
\left\langle h_c (p,\epsilon_{\lambda=0})\left|\bar{Q}^i_{\alpha} (z) [z,-z] Q^j_{\beta}(-z)\right|0\right\rangle  & = &
(\hat p \gamma_5)_{\beta \alpha}  \frac{f_{h}^{L}}{4}  \frac{\delta_{ij}}{3}
\int\limits _{-1}^{1}d\xi e^{i\xi(pz)}\phi_{h}^{L}(\xi;\mu),
\\
\left\langle h_c (p,\epsilon_{\lambda=\pm1})\left|\bar{Q}(z)\sigma_{\mu\nu}[z,-z]Q(-z)\right|0\right\rangle  & = &
(\hat p \hat \rho \gamma_5)_{\beta\alpha} \frac{f_{h}^{T}}{4}  \frac{\delta_{ij}}{3}
\int\limits _{-1}^{1}d\xi e^{i\xi(pz)}\phi_{h}^{T}(\xi;\mu)
\end{eqnarray*}

{\bf  for the $\chi_{c2}$-meson:}
\begin{eqnarray}
\left\langle \chi_{c2}(p,\epsilon_{\lambda=0})\left|\bar{Q}^i_{\alpha} (z) [z,-z] Q^j_{\beta}(-z)\right|0\right\rangle  & = &
(\hat p)_{\beta \alpha}  \frac{f_{\chi2}^{L}}{4} \frac{\delta_{ij}}{3}
\int\limits _{-1}^{1}d\xi e^{i\xi(pz)}\phi_{\chi2}^{L}(\xi;\mu),
\nonumber\\
\left\langle \chi_{c2}(p,\epsilon_{\lambda=\pm1})\left|\bar{Q}^i_{\alpha} (z) [z,-z] Q^j_{\beta}(-z)\right|0\right\rangle  & = &
M_\chi (\hat\rho \hat p)_{\beta \alpha} \frac{f_{\chi2}^{T}}{4} \frac{\delta_{ij}}{3}
\int\limits _{-1}^{1}d\xi e^{i\xi(pz)}\phi_{\chi2}^{T}(\xi;\mu).
\label{eq:phiChi2}
\end{eqnarray}
The factor $[z,-z]$, that
makes matrix elements (\ref{eq:phiChi2})
gauge invariant, is defined as
\begin{eqnarray}
[z, -z] = P \exp[i g \int_{-z}^z d x^{\mu} A_{\mu} (x) ].
\end{eqnarray}
In the above equations $p$ is the charmonium
momentum, $x$ and $\bar{x}$ are the momentum fractions of quark
and antiquark, $\xi=x-\bar{x}$, $\epsilon_{\mu}$ is the polarization
tensor for the $J/\psi$, $\chi_{c1}$ or $h_{c}$ mesons and the vector $\rho_{\mu}$
in relation (\ref{eq:phiChi2}) is defined according to
\begin{eqnarray*}
\rho_{\mu} & = & \frac{\epsilon_{\mu\nu}z^{\nu}}{(pz)},
\end{eqnarray*}
where $\epsilon_{\mu\nu}$ is the polarization tensor of the $\chi_{c2}$
meson. In practical applications it is useful to write the polarization of
the $\chi_2$ meson in terms of the polarization of two vector mesons. Thus, for instance,
the polarization tensor $\epsilon^{\mu \nu}$ of the transversely polarized $\chi_2$ meson
can be written as $\epsilon^{\mu \nu}_{\lambda=\pm 1} = (\epsilon_{\lambda=\pm1}^{\mu}\cdot \epsilon_{\lambda=0}^{\nu} +
\epsilon_{\lambda=0}^{\nu}\cdot \epsilon_{\lambda=\pm1}^{\mu})/\sqrt 2$~
($\epsilon^+_{\mu \nu} \epsilon^{\mu \nu}=1$).
If we further contract the polarization tensor $\epsilon_{\mu \nu}$ with lightlike four-vector $z$,
to the leading twist accuracy we will get
$\epsilon^{\mu \nu} z_{\nu} = \epsilon_{\lambda=\pm1}^{\mu} (pz)/(\sqrt 2 M_{\chi_2})$ or
$\rho^{\mu}=\epsilon_{\lambda=\pm1}^{\mu}/(\sqrt 2 M_{\chi_2})$. This form of the vector
$\rho$ can be used in the calculation with the leading twist accuracy.

It is not difficult to show  that the functions $\phi_{\eta}(\xi)$,
$\phi_{\psi}^{L,T}(\xi)$, $\phi_{\chi1}^{L}(\xi)$ and $\phi_{h}^{T}(\xi)$
are $\xi$-even. The normalization condition for these functions is
\begin{eqnarray}
\int\limits _{-1}^{1}\phi(\xi)d\xi & = & 1.\label{eq:normE}\end{eqnarray}
 The functions $\phi_{\chi0}^{L}(\xi)$,
$\phi_{\chi1}^{T}(\xi)$, $\phi_{h}^{L}(\xi)$ and $\phi_{\chi2}^{L,T}(\xi)$
 are $\xi$-odd and normalized according to \begin{eqnarray*}
\int\limits _{-1}^{1}\xi\phi(\xi)d\xi & = & 1.\end{eqnarray*}

\end{document}